# High Frame-Rate Mid-Infrared SPAD Camera

Ziv Abelson[1]*†, Daniel Beitner[1,2]†, Ziv Livne[2], Eyal Hollander[2], Moshe Cohen Erner[2], Edoardo Charbon[3] and Haim Suchowski[1]

[1] Condensed Matter Physics Department, School of Physics and Astronomy, Faculty of Exact Sciences and Center for Light-Matter Interaction, Tel-Aviv University, Tel-Aviv, Israel

[2] Spiral Photonics LTD, Herzliya, Israel

[3] Advanced Quantum Architecture Laboratory (AQUA), Ecole polytechnique fédérale de Lausanne (EPFL), 2002 Neuchâtel, Switzerland

† These authors contributed equally to this work

*Corresponding author. Email: zivabelson@mail.tau.ac.il, Tel:+972-50-2163544

| **Author** | **Email** |
|---|---|
| Ziv Abelson | zivabelson@mail.tau.ac.il |
| Daniel Beitner | beitner@mail.tau.ac.il |
| Ziv Livne | ziv@spiralphotonics.com |
| Eyal Hollander | eyal.e.hollander@gmail.com |
| Moshe Cohen Erner | moshe@spiralphotonics.com |
| Edoardo Charbon | edoardo.charbon@epfl.ch |
| Haim Suchowski | haimsu@tauex.tau.ac.il |

**Abstract:**

**Single-photon avalanche diode (SPAD) arrays have enhanced optical imaging capabilities by enabling a distinctive combination of photon-counting sensitivity, nanosecond-scale gating, and high frame-rate operation. These capabilities, however, have remained mostly confined to the visible and near-infrared, leaving the mid-infrared, the spectral region hosting the fundamental vibrational signatures of most molecules, largely devoid of these benefits. Here, we demonstrate the first mid-IR SPAD imaging system by integrating broadband adiabatic frequency upconversion with a 512 × 512 Silicon SPAD array. This architecture extends the SPAD functionality to the mid-IR, enabling room-temperature, low-noise, broadband photon-counting imaging. We achieve spectrally resolved mid-IR imaging at frame rates of up to 60,000 frames per second and capture nanosecond-scale laser-induced thermal dynamics via weak mid-IR blackbody emission, revealing spatial-temporal behavior predominantly indistinct to existing technologies. These results establish a potential platform for photon-resolved, ultrafast thermal and chemical imaging in a spectral range with more limited high-speed low-light detection capabilities.**

**Keywords:**, Nonlinear Optics, Upconversion Imaging, Infrared Imaging, Adiabatic Poling, Mid-IR SPAD, Single-Photon Imaging

## Introduction

Single-photon avalanche diode (SPAD) arrays have fundamentally enriched optical imaging by enabling true photon counting with nanosecond-scale exposure times, picosecond-scale delay capability, high frame rates, and digital noise-free readout [1,2] These capabilities have driven major advances in fluorescence lifetime imaging (FLIM) [3,4], time-of-flight sensing [5,6], ultrafast photon-resolved imaging [7], quantum communication [8], and photon-correlation measurements [9,10]. By shifting imaging from analog signal accumulation to photon-resolved detection, SPAD cameras have unlocked regimes typically elusive for conventional detectors. Although electron-multiplying CCD (EMCCD) and scientific CMOS (sCMOS) technologies provide high sensitivity, with feasible single photon detection capabilities and enable high speed low-light imaging. SPAD arrays extend the performance boundaries further through intrinsic photon counting, low noise, and nanosecond temporal gating at precise picosecond-scale timing delay.

Despite this progress, SPAD imaging remains largely confined to the visible and near-infrared spectral ranges [11], where silicon and InGaAs-based devices have reached technological maturity [12]. In contrast, the mid-infrared (mid-IR, 2-5 μm) region, home to the fundamental vibrational resonances of most molecules [13], relies on conventional detectors such as HgCdTe, which typically require cryogenic cooling, offer limited temporal resolution, and do not support photon-counting or ultrahigh frame rates [14,15]. This technological gap is particularly consequential because mid-IR light provides chemically specific contrast inaccessible at shorter wavelengths, underpinning applications in chemical identification [16,17], environmental monitoring [18,19], thermal science [20], astronomy [21,22], and medical diagnostics [23–25]. As a result, many of the capabilities that transformed visible-light imaging remain unavailable in one of the most information-rich spectral regions of the electromagnetic spectrum. Although recent advances have employed frequency upconversion to enable room-temperature mid-IR imaging that exploits the advantages of silicon detectors [26–29], with some works even implementing EMCCD based detection for imaging at the at the few-to-single photon level, these important advances have not implemented a genuine single photon detector or harnessed the performance benefits that a SPAD array possesses over EMCCD detection. In addition, the relatively low upconversion efficiency hinders SPAD based implementation since it drives the effective quantum efficiency levels even lower. Currently, mid-IR SPAD based single photon detection has only been demonstrated using an individual upconverted detector [30], while to the best of our knowledge, full-field mid-IR SPAD based single-photon imaging has not previously been realized.

Here, we bridge this gap by extending SPAD-based imaging into the mid-IR. By coupling a broadband adiabatic frequency upconversion platform to a silicon SPAD array, we realize a first high-frame-rate mid-IR SPAD imaging system that transfers the functionality of SPAD imaging, including photon counting, ultra-low noise, nanosecond-scale gating and picosecond-scale timing delay, into the $2 - 5\ \mu m$ band. Using a 512 × 512 SPAD array, we demonstrate broadband mid-IR imaging at frame rates up to 60,000 frames per second and exploit the system's temporal and spatial resolution to capture nanosecond-scale laser-induced heating and cooling dynamics via mid-IR thermal emission. The system operates at room temperature in a compact, small-footprint architecture with no moving parts or complex cooling apparatus. This work not only provides an integrative solution but also establishes a scalable platform for photon-resolved, ultrafast mid-infrared imaging, bringing to the mid-IR spectral range capabilities that have until now been mostly restricted to visible-light SPAD technologies.

# Results

## Mid-IR Single-Photon Imaging

Because scalable single-photon detector arrays in the mid-IR range remain technologically limited, alternative strategies are required. Silicon SPAD arrays represent the most advanced photon-counting imaging technology available, offering nanosecond scale gating, picosecond-scale delay timing, ultralow noise, and high frame rates. However, mid-IR photons lie well below the silicon bandgap and therefore cannot generate photoelectric charge carriers, rendering direct silicon-based detection impossible. Access to SPAD-based mid-IR imaging, therefore, requires photon-to-photon frequency conversion to translate the optical signal into the visible, where mature silicon SPAD arrays operate efficiently. Conventional nonlinear upconversion approaches are typically limited by narrow phase-matching bandwidths or require complex tuning and scanning to access broader spectra [31–33].

To overcome this limitation, we employ broadband frequency upconversion based on an adiabatically poled lithium niobate (APLN) crystal in an imaging configuration pumped by a Q-switched $1064\,nm$ laser (Fig. 1). In the APLN crystal, the poling period varies gradually along the direction of propagation, implementing adiabatic quasi-phase-matching that maintains efficient conversion over a broad spectral and angular range [34,35]. This adiabatic design enables robust frequency photon-to-photon conversion without fine-tuning, allowing a wide mid-IR bandwidth to be mapped into the visible-near-infrared domain in a single exposure. Throughout this work, illumination is provided by a filtered blackbody source spanning the $2-5\ \mu m$ spectral range. Residual pump light is rejected using dichroic filtering, and the upconverted image is recorded with the silicon SPAD array. The pump pulses are synchronized with SPAD gating to allow detection of the upconverted photons (See Fig. 1).

Due to the intrinsic dead time and binary operation of SPAD pixels, each frame (or number of accumulated gates) corresponds to a 1-bit image. While individual binary frames contain sparse photon events, synchronized acquisition enables summation of multiple frames, yielding high-sensitivity images with ultralow noise. This additive photon-counting scheme can be extended beyond static scenes to repetitive dynamic processes by synchronizing acquisition to external events with picosecond-scale delay precision as illustrated in Fig. 1. Using this approach, we capture images of a USAF 1951 resolution test target in the mid-IR range (Fig. 2). As shown in Fig. 2a, individual SPAD frames contain sparse binary photon-detection events and provide minimal spatial information. Summation of increasing numbers of frames (N = 1–1000) progressively reconstructs the image, revealing fine spatial features with increasing contrast as photon statistics accumulate. Beyond sensitivity and spatial resolution, this imaging arrangement also enables a multispectral imaging capability. Spectral sectioning of the mid-IR signal is implemented by inserting narrowband filters in the visible after upconversion, without modifying the nonlinear conversion stage. Because the process obeys energy conservation, a narrow visible-band filter corresponds to a broader spectral window in the mid-IR range. For example, a $10\,nm$ FWHM filter centered near $800\,nm$ selects an effective mid-IR bandwidth of approximately $160\,nm$ around $3.5\ \mu m$. Fig. 2b shows mid-IR images acquired at different spectral sections across the $2-5\ \mu m$ band. With optimized gating, the mid-IR SPAD imaging system resolves features as small as 35 $\mu m$ throughout this range. Variations in magnification and field of view between spectral sections arise from momentum-conservation-induced dispersion [36] and the wavelength-dependent angular acceptance of the adiabatically poled crystal. Further narrowing the spectral filtering improves image contrast at the expense of optical throughput. Resulting wavelength dependent features introduced by the crystal can be mitigated through straightforward image processing, described in the Supplementary Information section.

When examining the functionality of a SPAD array, several key parameters are important to establish and compare the performance of the mid-IR SPAD imaging system. The end-to-end detection efficiency of the system is dependent on the upconversion efficiency and the silicon SPAD photon detection efficiency, which are both wavelength dependent. Numerical calculation and experimental measurements, which are described in depth in the Supplementary Information section, lead to an estimated efficiency of the order of 0.1% throughout the $2-5\ \mu m$ spectral range (see Section S3.1). In addition, dark count analysis reveals a median dark-count-rate (DCR) of $\sim 2000\ cps$ for spectrally sectioned regions and a DCR of $\sim 125{,}000\ cps$ for full spectrum images (see Section S3.2 of the Supplementary Information). These factors lead to a substantially higher signal-to-noise (SNR) ratio compared to standard SPAD cameras (see Section S5).

A defining advantage of SPAD-based imaging is the achievable frame rate, which remains limited in conventional infrared cameras, particularly in compact, room-temperature systems. As shown in Fig. 3, we capture dynamic mid-IR scenes at high frame rate, including an optical chopper rotating at a chopping frequency of $900\ Hz$ and a butane burner that was positioned 15 meters from the adiabatic mid-IR SPAD imaging system. These measurements demonstrate the ability to resolve rapid temporal dynamics and evolving thermal features in the mid-IR at frame rates exceeding those of standard infrared imaging technologies.

Fig. 3b and 3c show time-resolved images of the rotating chopper acquired as 1-bit and 8-bit frames, respectively. The SPAD array supports frame rates up to 100,000 frames per second (fps) in 1-bit mode, but in practice, the achievable rate is governed by the available photon flux. For the butane burner (Fig. 3d-f), images acquired at lower frame rates exhibit motion-induced blurring due to flame dynamics. Increasing the frame rate reveals progressively sharper spatial features (Fig. 3e), while at the highest frame rates only the most intense emitting regions remain visible (Fig. 3f), reflecting photon-limited acquisition. Together, these results highlight the mid-IR SPAD platform's ability to capture fast-evolving thermal phenomena over both laboratory and field-relevant distances.

## Capturing Nanosecond Dynamics in the Mid-IR

To demonstrate the temporal capabilities of the mid-IR SPAD imaging system, we showcase nanosecond laser-induced heating in various metals. As shown in Fig. **4**a, a synchronized $1064\ nm$ Nd:YAG laser ($1\ mJ$ maximum pulse energy) is focused onto the metal surface to a spot diameter of approximately $200-400\ \mu m$, with the fluence maintained well below the ablation threshold. The sample is positioned in front of the system such that the excitation beam impinges perpendicularly at a grazing angle. The heating laser, upconversion pump, and SPAD gating are synchronized and operated at a fixed repetition rate. By incrementally shifting the relative delay in steps of $5\ ns$, we reconstruct the full spatiotemporal evolution of the mid-IR blackbody emission ($2-5\ \mu m$) following each excitation pulse. Fig. **4**c shows representative time-resolved images and spatially integrated dynamics.

Direct imaging of nanosecond-scale mid-IR thermal emission is intrinsically challenging because the emitted photon flux is low and conventional detectors require longer integration times, increasing noise and limiting temporal resolution. The photon-counting sensitivity and synchronized gating of the mid-IR SPAD platform enable access to this regime with spatial resolution. To our knowledge, this represents the first spatially resolved nanosecond-scale mid-IR thermal emission imaging enabled by full-field photon counting.

The measurements reveal cooling dynamics that extend well beyond the $\sim 8\ ns$ duration of the excitation pulse. The spatial heating profile follows the Gaussian intensity distribution of the laser beam, while the temporal response exhibits two distinct contributions. The first is rapid cooling of the metal following each pulse, and the second is a weaker steady-state temperature resulting from repetitive pulses' excitation. The complete spatial-temporal evolution is readily resolved, and wavelength-dependent sectioning provides additional insight into the thermal dynamics. Fig. **4**d illustrates the time-resolved spatiotemporal evolution of the thermal emission, across horizontal and vertical cross-sections, providing an additional perspective on the thermal profile's spatial diffusion and decay.

The cooling dynamics vary markedly across different metals (Fig. **4**e), reflecting differences in specific heat, thermal conductivity, and heat diffusion length. This sensitivity suggests a route toward material discrimination based on time-resolved mid-IR thermal signatures and may be extended to time-resolved absorption or emissivity studies. Additionally, an example of spectral sectioning of the thermal dynamics is provided in supplementary information Fig S2.

Conventional characterization of fast pulsed laser heating dynamics has been typically performed by probing reflectivity changes in the visible or near-infrared [37,38], characterizing plasma emission [39] or by pulsed photothermal radiometry [40,41]. Since direct IR emission is extremely weak and short lived, these methods rely on signal integration to obtain measurable data. In contrast, the present approach provides the entire images of mid-IR thermal emission at nanosecond resolution, providing not only the temporal response but also complementary spatial insights into energy deposition and heat transport processes.

## Discussion

We extend SPAD-based imaging into the mid-IR by combining adiabatic frequency upconversion with large-format silicon SPAD arrays, enabling photon-counting, high-frame-rate imaging across the 2–5 μm band at room temperature. This platform bridges a longstanding technological gap by bringing a distinctive combination of nanosecond-scale gating, picosecond-scale delay resolution, ultralow noise, high binary frame-rates and full-field single-photon sensitivity to a spectral region in which such capabilities were generally limited. In doing so, it establishes a scalable approach for high-speed, photon-resolved mid-IR imaging without cryogenic cooling and any moving components.

The present architecture enables mid-IR imaging without cryogenic cooling of the detector array and without mechanically scanned components. While the current implementation already extends SPAD-based imaging into the mid-IR, further improvements in end-to-end detection efficiency and background suppression will be important for practical deployment. At present, the system exhibits an overall detection efficiency of the order of $\sim 0.1\%$ and a relatively high equivalent dark-count/background level (see Supplementary Information). Reaching system detection efficiencies in the range of approximately $1\text{–}10\%$ would substantially improve sensitivity and reduce the integration time required for photon-limited measurements.

One direct route toward this goal is to increase the per-pulse upconversion efficiency. The present architecture combines an adiabatically poled lithium niobate crystal with a nanosecond Q-switched pump, providing high peak power within a short temporal window that is naturally compatible with gated SPAD detection. Higher peak pump power can further increase the nonlinear conversion efficiency, while synchronization of the pump pulse with the SPAD gate ensures that the upconverted signal is generated during the detector exposure window. A pump source specifically optimized for this mode of operation could therefore be designed to maintain the required peak power density over the relevant gate duration, maximizing temporal overlap between conversion and detection.

The combination of Q-switched pumping and adiabatic poling is particularly attractive in this context because it provides useful per-pulse conversion efficiency while preserving broadband spectral and angular acceptance. This architecture therefore offers a clear path toward higher-efficiency implementations. Practical optimization will nevertheless require balancing pump peak power against system footprint and complexity, while remaining below the optical-damage threshold of the APLN crystal.

A second limitation of the current implementation arises from the wavelength dependence of the silicon SPAD photon-detection efficiency (PDE). The detector PDE peaks near $520\ nm$, while the upconverted wavelengths corresponding to part of the present mid-IR operating range fall in a region where the PDE is lower. Future implementations could therefore improve the spectral matching between the nonlinear-conversion stage and the detector in two complementary ways: by selecting or engineering silicon SPAD detectors with enhanced PDE at the relevant upconverted wavelengths, and by adjusting the pump wavelength so that the converted mid-IR signal is shifted toward the region of maximum SPAD sensitivity. This additional degree of freedom provides a practical route to increasing the end-to-end detection efficiency without changing the targeted mid-IR spectral band.

Background suppression provides an additional route toward improved sensitivity. In the present system, residual pump and second-harmonic light contribute significantly to the equivalent dark-count rate. More selective spectral filtering and improved rejection of these parasitic components could therefore substantially improve the signal-to-background ratio. Because the SPAD gate can be synchronized to the pump pulse, background photons arriving outside the nonlinear-conversion window are intrinsically rejected, providing an additional temporal discrimination mechanism beyond spectral filtering. The measurements presented here already show that spectral filtering reduces the background count rate and improves image quality, although this comes at the expense of discarding part of the broadband signal. Future optimization of both the spectral filtering and synchronized gating conditions should therefore allow the signal throughput and background rejection to be tailored to the requirements of the specific application.

The nanosecond-resolved thermal-emission measurements also illustrate a distinct application space for the system. Nanosecond and ultrafast lasers are widely used in machining, fabrication, drilling, welding, annealing, and other laser-based manufacturing processes, where transient energy deposition and heat flow can strongly influence process quality and material response. Spatially and temporally resolved mid-IR measurements could provide direct access to the evolution of the heated region, cooling dynamics, and spatial heat transport, and may therefore be useful for process monitoring, optimization and failure analysis.

More broadly, the ability to measure spatially resolved mid-IR emission on nanosecond timescales demonstrates the potential of this architecture for studying fast energy-deposition and heat-transport processes. Beyond the demonstrations presented here, synchronized delay-scanning measurements could be applied to lifetime-resolved dynamics, high-speed vibrational imaging, and non-equilibrium thermal and emissive processes requiring combined spatial, spectral, and temporal information. In this sense, extending gated SPAD detection into the mid-IR provides a complementary platform for experiments that benefit from photon-counting sensitivity, fast acquisition, broadband spectral access, and precisely synchronized temporal gating.

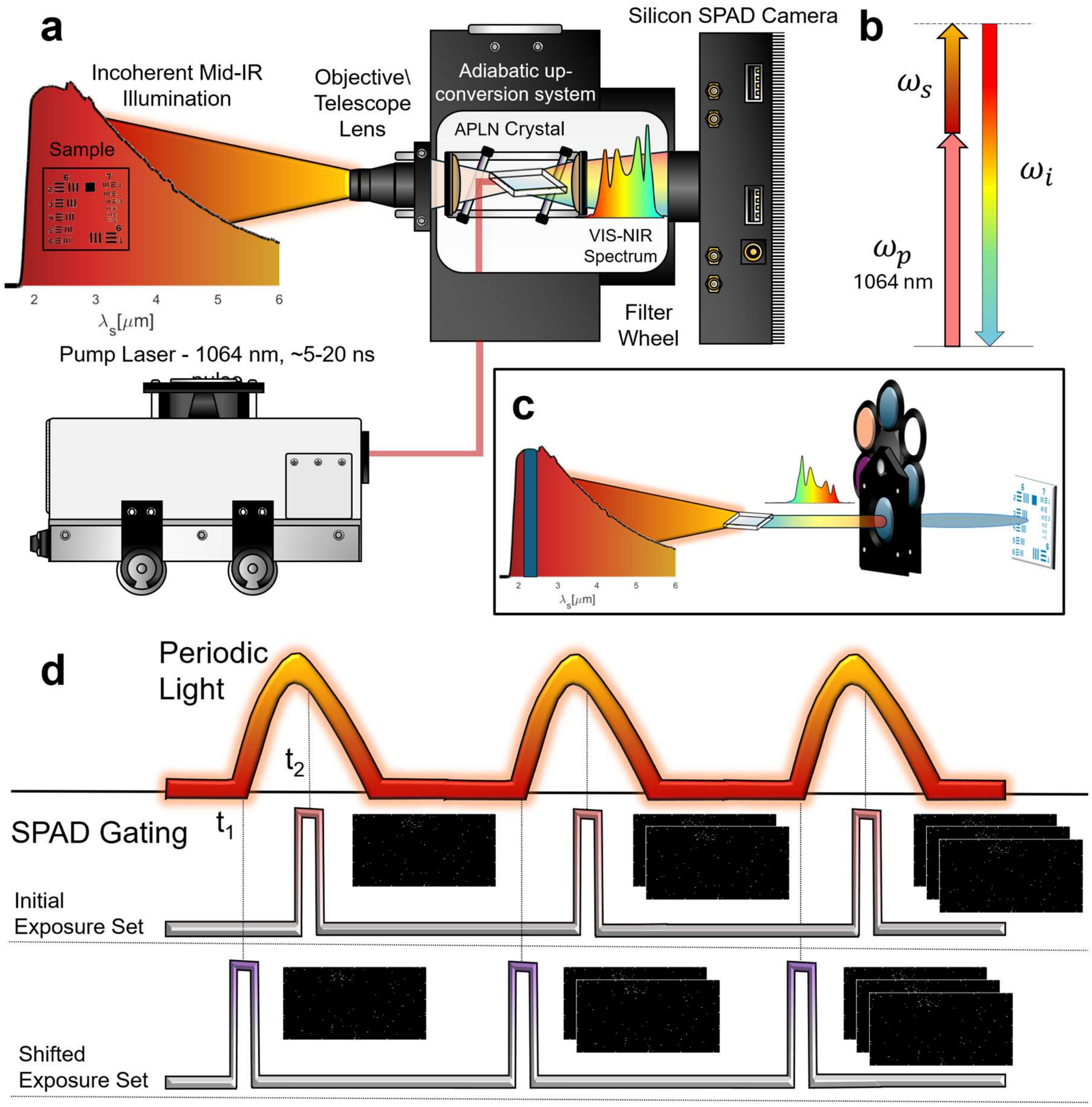

**Fig. 1. Principles of mid-IR SPAD camera operation. a** Schematic of the adiabatic upconversion imaging configuration. Blackbody radiation filtered to the $\mathbf{2-5}\ \boldsymbol{\mu m}$ band is collected by an objective lens and relayed in a 4f system, with the APLN crystal positioned at the Fourier plane. The crystal, providing broadband spectral and angular acceptance, is pumped by a high peak-power Nd-YAG laser (1064 nm, $\mathbf{5-20}\ \boldsymbol{ns}$ pulse, $\mathbf{0.1-1}\ \boldsymbol{mJ}$, depending on repetition rate). The mid-IR image is upconverted to the $\mathbf{680-880}\ \boldsymbol{nm}$ visible–near-infrared range, where spectral filtering enables selection of specific mid-IR bands. **b** Energy conservation diagram illustrating the wavelength relationship in the upconversion process. **c** Illustration of the spectral sectioning. Since the upconversion is governed by energy conservation, filtering the visible output selects the corresponding mid-IR spectral segment. **d** Illustration of SPAD-based time-resolved acquisition for fast periodic phenomena (Gated Mode). Binary frames are accumulated at a fixed delay to form a high-sensitivity image. The delay is then gradually incremented to reconstruct the temporal evolution of a periodic event.

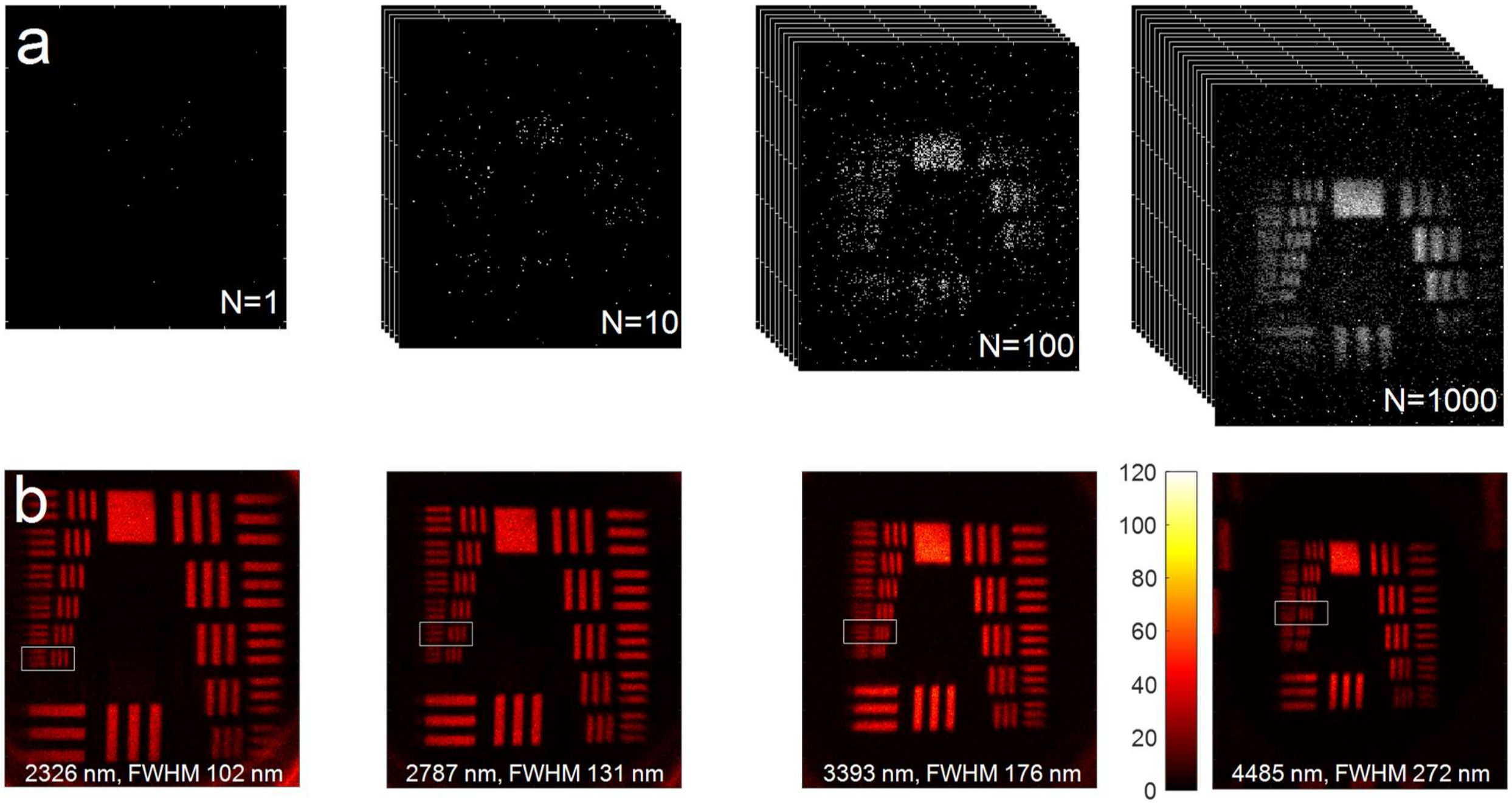

**Fig. 2. Resolution and spectral sectioning of mid-IR images. a** Progressive reconstruction of a mid-IR image through summation of binary single-photon frames acquired after frequency upconversion using a visible spectral filter centered at $\mathbf{850}\ \boldsymbol{nm}$ ($\boldsymbol{FWHM}\ \mathbf{10}\ \boldsymbol{nm}$). For a pump wavelength of $\mathbf{1064}\ \boldsymbol{nm}$, this corresponds to a mid-IR spectral band centered at $\mathbf{4.23}$ μ$\boldsymbol{m}$ with an effective bandwidth of $\mathbf{250}\ \boldsymbol{nm}$), determined by the energy-conservation relation $\boldsymbol{\lambda}_{\mathbf{IR}}^{-1} = \boldsymbol{\lambda}_{\mathbf{VIS}}^{-1} - \boldsymbol{\lambda}_{\mathbf{pump}}^{-1}$. Individual frames (N = 1) contain sparse photon-detection events, while summation over increasing numbers of frames (N = 10, 100, 1000) progressively reconstructs the image, revealing increasing contrast and spatial detail. 1-bit Images were captured for $\mathbf{50}\ \boldsymbol{ns}$ SPAD integration time and a frequency of 60 kHz, with each gate generating a 1-bit image. Images were captured in Intensity mode. **b** Mid-IR images of a USAF1951 resolution test target acquired at different spectral sections across the $\mathbf{2-5}$ μ$\boldsymbol{m}$ band, as indicated. 10-bit images were captured for $\mathbf{20}\ \boldsymbol{ns}$ SPAD gating, a frequency of 5 kHz, an integration time of 1.2 s, corresponding to 6000 gating events and a total exposure time of $\mathbf{120}\ \boldsymbol{\mu s}$, with a signal-to-noise-ratio ranging between 35-60 for each of the images. The colorbar indicates photon counts for each pixel for all four images. For the $\mathbf{2.3}\ \boldsymbol{\mu m}$ image, we observe the smallest features of element 6 of group 3 (marked on the image), which correspond to a spatial resolution of approximately $\mathbf{35}\ \boldsymbol{\mu m}$. Provided this is the final resolution group of the target, the system resolution at this wavelength is expected to be finer. For longer wavelengths of $\mathbf{2.8-4.5}\ \boldsymbol{\mu m}$, only element 5-4 in group 3 are discernible and this indicates a spatial resolution of $\mathbf{40-45}\ \boldsymbol{\mu m}$ for each spectral region. Variations in magnification and field of view between spectral sections arise from momentum-conservation-induced dispersion and the wavelength-dependent angular acceptance of the adiabatically poled lithium niobate (APLN) crystal. Further analysis of the optical characteristics is presented in the Supplementary Information section.

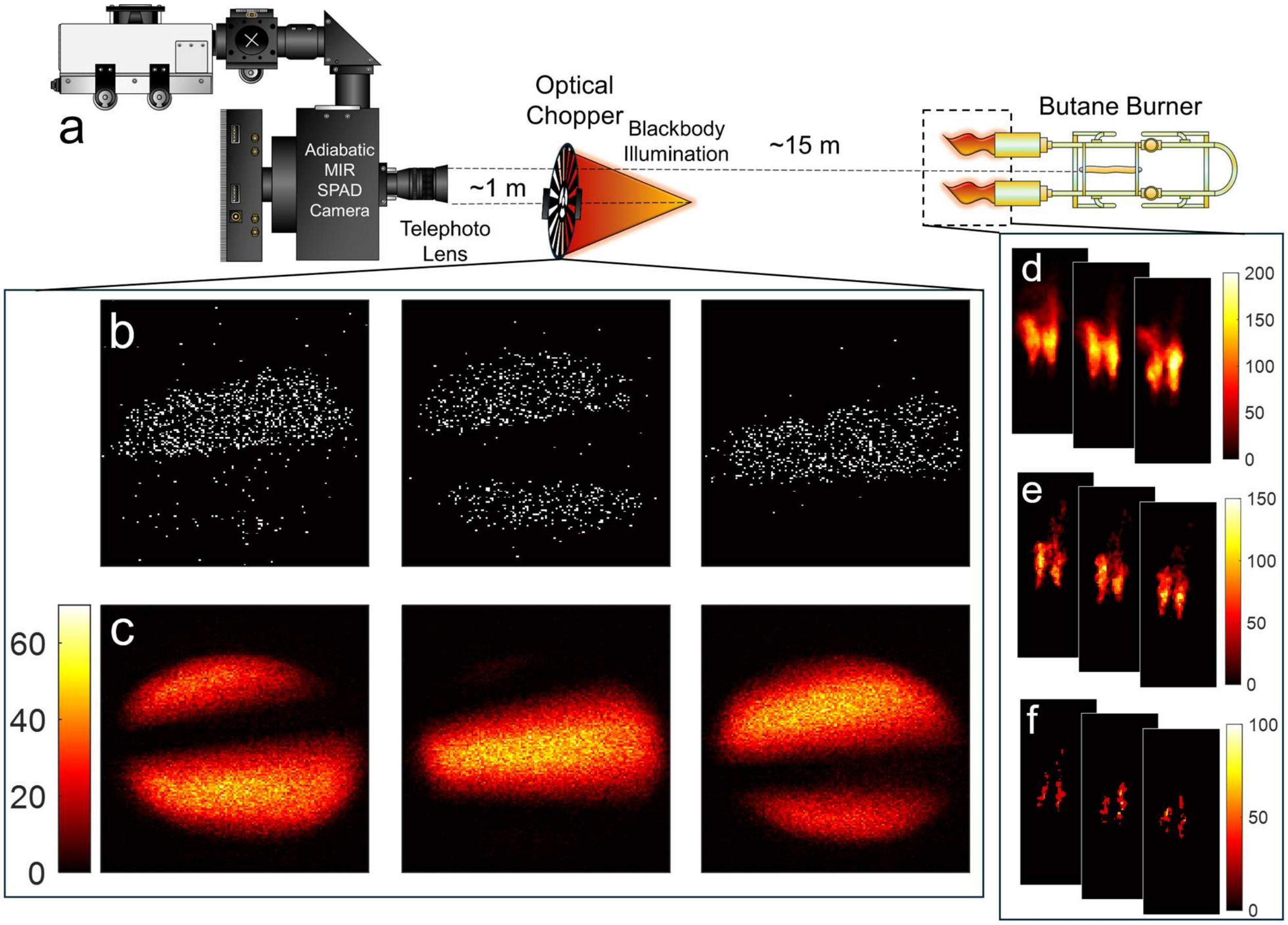

**Fig. 3. High Frame Rate mid-IR SPAD Imaging. a** Experimental configuration for high-frame rate mid-IR imaging. The mid-IR SPAD imaging system was equipped with a telephoto lens, and dynamic scenes were captured under filtered blackbody illumination (optical chopper) and under filtered natural emission (dual-head butane burner), providing the full mid-IR spectrum. **b** 1-bit images of an optical chopper rotating at a chopping frequency of 900 Hz acquired with $\mathbf{50}\ \boldsymbol{ns}$ integration time and at a repetition rate of 60 kHz. Images were captured in Intensity mode. **c** 8-bit images of the optical chopper rotating at a chopping frequency of 300 Hz acquired with an $\mathbf{11}\ \boldsymbol{\mu s}$ integration time (Intensity mode), with a typical $\boldsymbol{SNR} \cong \mathbf{10}$. **d-f** Mid-IR images of a dual-head butane burner located $\mathbf{15}\ \boldsymbol{m}$ from the imaging system, acquired with $\mathbf{50}\ \boldsymbol{ns}$ gating time, at a repetition rate of 60 kHz and a typical $\boldsymbol{SNR} \cong \mathbf{4-45}$. **d** 8-bit images with frame rates of $\mathbf{25}\ \boldsymbol{fps}$, an integration time of $\mathbf{40}\ \boldsymbol{ms}$, corresponding to 2400 gating events and a total exposure time of $\mathbf{120}\ \boldsymbol{\mu s}$. **e** 6-bit images with frame rates of $\mathbf{250}\ \boldsymbol{fps}$, an integration time of $\mathbf{4}\ \boldsymbol{ms}$, corresponding to 240 gating events and a total exposure time of $\mathbf{12}\ \boldsymbol{\mu s}$. **f** 6-bit images with frame rates of $\mathbf{1000}\ \boldsymbol{fps}$, integration time of $\mathbf{1}\ \boldsymbol{ms}$, corresponding to 60 gating events and a total exposure time of $\mathbf{3}\ \boldsymbol{\mu s}$. The colorbars indicate photon counts for each pixel.

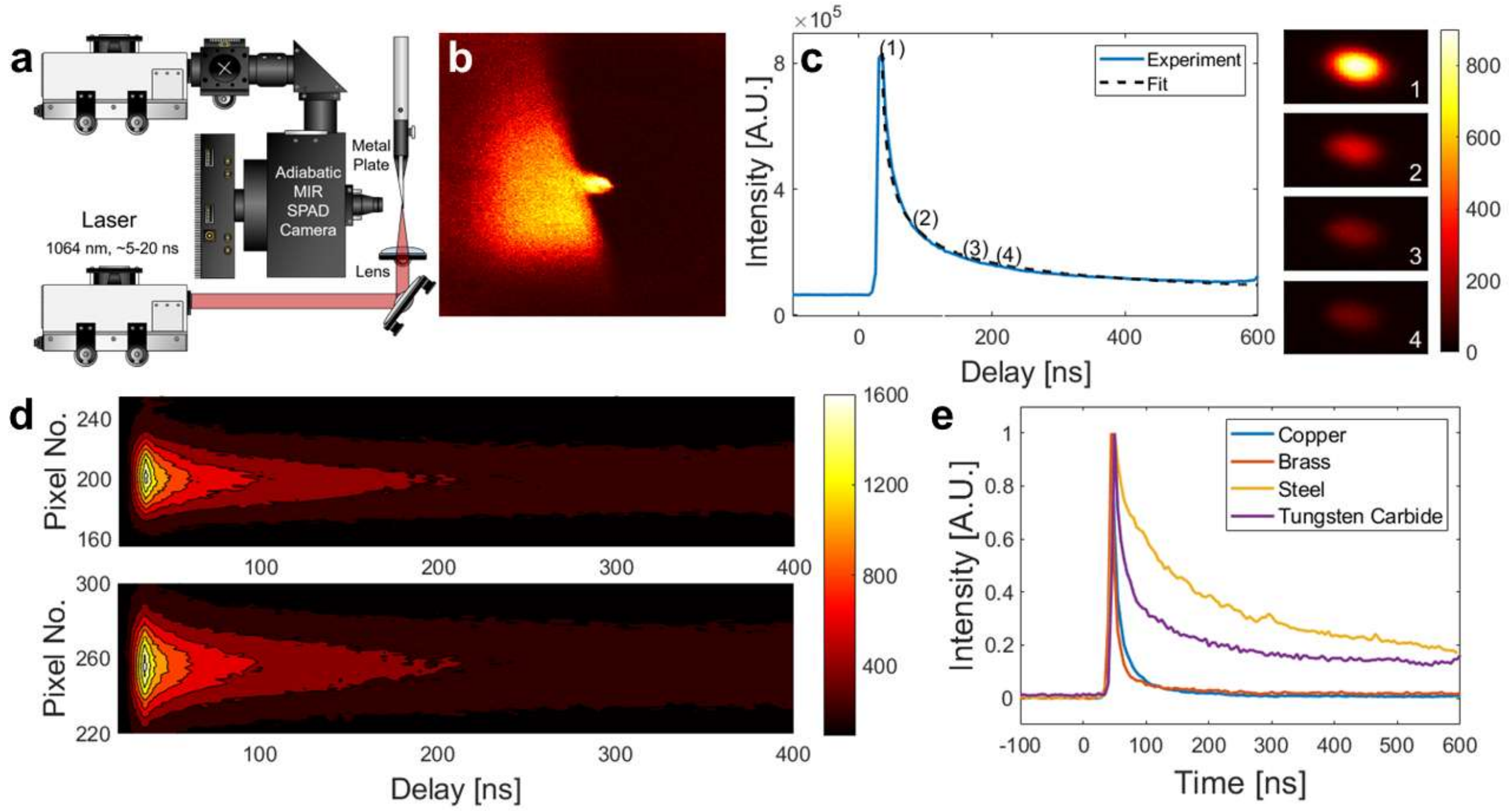

**Fig. 4. Nanosecond-scale thermal dynamics in the mid-IR. a** Experimental configuration for time-resolved (ns-scale) mid-IR imaging of laser-induced heating in metals. A thin metal plate is positioned in front of the mid-IR SPAD imaging system, while a synchronized $\mathbf{1064}\,\boldsymbol{nm}$ laser pulse ($\mathbf{5-20}\,\boldsymbol{ns}$, $\mathbf{0.1-1}\boldsymbol{mJ}$, $\mathbf{5}\,\boldsymbol{kHz}$) is focused onto the surface at a grazing angle. **b** Representative mid-IR image showing the spatial profile of the laser-heated region immediately following excitation. This image was captured using external blackbody illumination. **c** Temporal cooling dynamics following a single laser pulse, obtained by integrating the mid-IR signal over a region-of-interest (ROI). Numbered markers correspond to the time delayed mid-IR images shown to the right. The cooling profile corresponds to the typical temporal shape obtained by pulsed photothermal radiometry measurements and the model suggested by Tam [41]. **d** Time-resolved spatiotemporal evolution of the thermal emission, shown as X- and Y- axis cross-sections through the heated region (top and bottom, respectively), revealing the diffusion and decay of the thermal profile from nanosecond to sub-microsecond timescales. **e** Peak normalized cooling dynamics for different metals, illustrating distinct temporal responses arising from variations in thermal properties such as specific heat and thermal conductivity (see Supplementary Information Fig. S1 for examples of spatiotemporal evolution for different materials). For **b-e,** full spectrum images were captured for $\mathbf{20}\,\boldsymbol{ns}$ SPAD gating, a frequency of 5 kHz, an integration time of 1.2 s, corresponding to 6000 gating events and a total exposure time of $\mathbf{120}\,\boldsymbol{\mu s}$, displaying a typical $\boldsymbol{SNR} \cong \mathbf{60}$. Heating laser fluence was between $\mathbf{0.2-0.7}\,\boldsymbol{J/cm^2}$. Apart from **b,** all images were captured without external blackbody illumination. The colorbars indicate photon counts for each pixel.

# Materials and Methods

## Nonlinear Crystal Design

Throughout this work, various adiabatic poling designs have been implemented and tested with different crystals. Initial designs were manufactured with dimensions of 2 mm×3 mm×3 mm. To further improve efficiency, the more recently designed crystals were fabricated with increased length (2 mm × 3 mm × 5 mm). For each series, designs were implemented to either optimize bandwidth, spectral response or material identification. The quasi-phase matching period ranges between a minimum value of $14.8\ \mu m$ and a maximum value of $31\ \mu m$ for all the designs and series presented.

The adiabatic design is formulated according to the evolution of the phase mismatch in k-space rather than by imposing a simple linear chirp of the poling period in real space. This approach enables a controlled trajectory through the phase-matching condition, providing a broad and smooth spectral response and increased robustness to variations in wavelength and other interaction parameters.
The crystals are designed for type-0 nonlinear interaction, with the mid-IR signal, pump, and upconverted idler all polarized along the extraordinary axis of the lithium niobate crystal.

## Experimental Setup – Adiabatic Mid-IR Imaging System

Here, the principles of operation of the SPAD-based adiabatic mid-IR imaging system are presented. A schematic which illustrates the details of the system is provided in the main body of the work. In each experiment, the light is collected either by a lens pair (for magnification) or telephoto lens and directed into a 4f system, with $f_{MIR} = 12.7\ mm$ and $f_{VIS} = 50mm$. The center of the APLN crystal is placed at the Fourier plane of the 4f system. The mid-IR light is generated either by a thermal process such as laser heating or combustion, or the sample under examination is illuminated by an external source, in this case a cavity-stabilized blackbody radiation source (CI systems SR-20-32) heated to $1000^{o}C$. A Germanium window with $2 - 5\ \mu m$ anti-reflective coating is set before the 4f system and serves as a long-pass filter to ensure the suppression of wavelengths below $1.7\ \mu m$ to noise levels.

A $1064nm$, $\sim 0.1 - 1\ mJ$, $5 - 20\ ns$ pump beam operating at $5 - 100\ kHz$ (BrightSolutions Onda 1064) is inserted into and aligned with the optical path. The pump beam was collimated to a diameter of between $1 - 2\ mm$ to fit the size of the adiabatic crystal and optimized for either higher peak power or better imaging resolution. The SFG process takes place in the crystal and converts the mid-IR signal to the corresponding visible-NIR spectrum. Before exiting the 4f arrangement, the light is then carefully filtered by a dichroic mirror designed to reflect the pump beam, it's second harmonic and any mid-IR illumination away from the optical path. Only light in the range of $660\ -\ 1000\ nm$ is transmitted beyond this stage.

Finally, A filter wheel has been installed before the detector to easily separate the multi-spectral image into narrower regions and can be conveniently adapted to enable the division of the spectrum into different segments per requirement. The filter wheel generally employs bandpass filters with a center wavelength in the $690\ -\ 900\ nm$ range and a typical FWHM of $10\ nm$. A silicon-based SPAD camera (Pi-Imaging SPAD512) serves as the detector of the incident light. Thus, a broadband illumination spectrum can be converted concurrently to an equivalent spectrum of images at shorter wavelengths. It is possible to obtain either a wide range multispectral image or a series of narrower range images in the $2-5\ \mu m$ region.

Regarding integration between the different components of the system, it is crucial to synchronize between the SPAD gating and the pump pulse to obtain upconversion. Moreover, for any experiment conducted with a pulsed or repetitive source, synchronization is possible and should be conducted. For the laser heating experiments that were performed, both the SPAD camera and the two lasers were synchronized together with an initial fixed delay, to compensate for the latency of each component. To temporally examine the phenomenon, the SPAD gating and pump pulse delay were shifted in relation to the heating beam.

**SPAD Camera Functionality**

It is important to note that the capability of the silicon SPAD camera supports 1-bit image capture of up to $100{,}000\ fps$, with the array readout time being the limiting factor. For higher dynamic ranges, the frame rate drops drastically, e.g. $5{,}000\ fps$ for 4-bit images and $400\ fps$ for 8-bit images, while enabling longer integration times. This serves as a crucial trade-off between image quality and frame rate of the mid-IR imaging system, which is especially relevant for asynchronous phenomena and should essentially be considered with the specific application in mind. The SPAD camera supports two different operation modes – Intensity Mode, which is a photon counting mode that is active throughout each defined exposure period, similar to standard camera operation with image capture at the configured frame rate. Gated Mode, on the other hand, applies a precision-controlled electronic "gate" to allow short synchronized and repeated exposures.

For Gated Mode, higher dynamic range is achieved in the following manner - the SPAD camera generates gated events (i.e. single exposure events) according to the defined gate width $t_{gate}$ at the selected repetition rate $f$. Thus, the number of gated events is the product of the frequency and the integration time $t_{int}$, $N_{gates} = t_{int} \cdot f$ and the estimated total exposure time $t_{total\ exposure} = N_{gates} \cdot t_{gate}$. To produce an n-bit image, a series of $2^n - 1$, 1-bit images is generated. This is achieved by summing up groups of $\cong \frac{N_{gates}}{2^n-1}$ gates and performing readout. The separate $2^n - 1$, 1-bit images are then accumulated to construct an n-bit image.

**Pump Laser Parameters**

In Table 1 we provide the main parameters that were measured for pump laser operation. Note that there is a strong dependence between the operating frequency and laser power. Thus, in many instances there is justification for operating at lower repetitions rates to obtain better conversion efficiency. The power density was calculated for a beam diameter of 1.8 mm, which is the typical beam diameter used for the majority of measurements performed in this work. The pump laser polarization is linear and set on the vertical axis. The power stability of the laser, after initial warmup, is below $< 2\%$.

The typical spectral width of the pump laser that has been employed in the system is $< 150\ pm$. The induced spectral broadening of the upconverted light only due to the linewidth of the pump is sub-nanometer for all values in the $2-5\ \mu m$ range. This broadening is negligent compared to the spectral width of any of the bandgap filters that have been used to spectrally segment the upconverted signal, which possesses a typical FWHM of $10\ nm$.

Given the dependence between the repetition rate of the laser and the pulse power, there is a possible trade-off between the number of gating events and the per-pulse efficiency for each event. It has been confirmed that a frequency of $60\ kHz$ is the optimal operational frequency in terms of system operation. Yet, at $5\ kHz$, the peak pulse power is maximal and in correlation the conversion efficiency is the highest. In low light scenarios, such as the laser heating experiment this can be cardinal and possess a noticeable effect on the SNR.

| Repetition Rate [kHz] | Pulse Power [mJ] | Average Power [W] | Pulse Width [ns] | Peak Power [kW] | Peak Power Density $[MW/cm^2]$ |
|---|---|---|---|---|---|
| 5 | 0.8 | 4.1 | 7.8 | 105.1 | 8.3 |
| 20 | 0.6 | 11.2 | 8.1 | 69.5 | 5.5 |
| 40 | 0.3 | 13.5 | 8.9 | 38.1 | 3.0 |
| 60 | 0.2 | 14.1 | 10.4 | 22.5 | 1.8 |
| 80 | 0.2 | 14.3 | 13.8 | 13.0 | 1.0 |
| 100 | 0.1 | 14.3 | 20.7 | 6.9 | 0.5 |

**Table 1: Measured pump laser parameters**

## Data availability

The data supporting the findings of this study are available from the corresponding author upon reasonable request.

## Conflict of interest

All authors declare no competing interests.

## Contributions

Z.A., D.B., E.C., and H.S. conceptualized and designed the study. Z.A., D.B., and M.C.E. performed the experimental characterization, measurements, and data analysis. E.H. and H.S. designed the APLN crystals. H.S. and Z.L. supervised and managed the project. Z.A. and H.S. wrote the manuscript with input from all authors. All authors contributed to the discussion and interpretation of the results, reviewed and revised the manuscript, and approved its final version.